# Pore structure and electrochemical properties of CNT-based electrodes studied by *in situ* small/wide angle X-ray scattering[†]

Cleis Santos,[a] Evgeny Senokos,[a,b,c] Juan Carlos Fernández-Toribio,[a,b] Álvaro Ridruejo,[b] Rebeca Marcilla[c] and Juan José Vilatela [a*]

Macroscopic ensembles of nanocarbons, such as fibres of carbon nanotubes (CNT), are characterised by a complex hierarchical structure combining coherent crystalline regions with a large porosity arising from imperfect packing of the large rigid building blocks. Such structure is at the centre of a wide range of charge storage and transfer processes when CNT fibres are used as electrodes and/or current collectors. This work introduces a method based on wide and small-angle X-ray scattering (WAXS/SAXS) to obtain structural descriptors of CNT fibres and which enables *in situ* characterisation during electrochemical processes. It enables accurate determination of parameters such as specific surface area, average pore size and average bundle size from SAXS data after correction for scattering from density fluctuations arising from imperfect packing of graphitic planes. *In situ* and *ex situ* WAXS/SAXS measurements during electrochemical swelling of CNT fibre electrodes in ionic liquid provide continuous monitoring of the increase in effective surface area caused by electrostatic separation of CNT bundles in remarkable agreement with capacitance changes measured independently. Relative contributions from quantum and Helmholtz capacitance to total capacitance remaining fairly constant. The WAXS/SAXS analysis is demonstrated for fibres of either multi- and single-walled CNTs, and is expected to be generally applicable to *operando* studies on nanocarbon-based electrodes used in batteries, actuators and other applications.

Carbon-based materials are present in virtually all energy-storing devices. They can combine light weight, high electrochemical stability, electrical conductivity and large surface area. Their basic units are small crystalline ("graphitic") domains of hexagonal $sp^2$ conjugated carbon atoms, together with defects such as topological (e.g. pentagons), edges, heteroatoms. [1] However, in traditional graphitic materials there is an inherent compromise between crystal size and porosity, which implies a trade-off between transport properties and electrochemical stability, and specific surface area (SSA). Highly graphitic carbons used in batteries and fuels cells, for example, have a crystal size of around 20-100 $nm$, conductivities as high as $2 \times 10^4\ Scm^{-1}$, but little porosity. [2–4] Alternatively, defects are used to introduce curvature and prevent stacking of graphitic layers, which increases SSA, a driver in electrical double-layer (EDL) supercapacitors. A porous activated carbon with an average crystal size of 6-18 $nm$, has a very high SSA of 900-2500 $m^2g^{-1}$, but electrical conductivity of 25-75 $Scm^{-1}$. [5–8]

This inherent compromise between crystal size and porosity does not apply to carbon nanotubes (CNTs) or other nanocarbons (graphene, etc). Assembled as macroscopic fabrics or yarns, the porosity of the ensemble arises from imperfect packing of the nanocarbon "molecules", and is decoupled from the degree of perfection of the graphitic molecules, which depends exclusively on the method of synthesis and processing. In the case of macroscopic fibres/yarns of CNTs, for example, this translates into specific surface areas around 250 $m^2g^{-1}$ and crystal size as high as 2 $\mu m$. [9,10] In terms of bulk properties, this implies combined capacitance of 10-78 $Fg^{-1}$ with tensile properties approaching those of carbon fibres and mass-normalised electrical conductivity above some metals. [11,12] Furthermore, yarns made up of highly graphitised few-layer CNTs have been also shown to preserve low-dimensional properties, such as accessible quantum (chemical) capacitance due to their low joint density of electronic sates. [9]

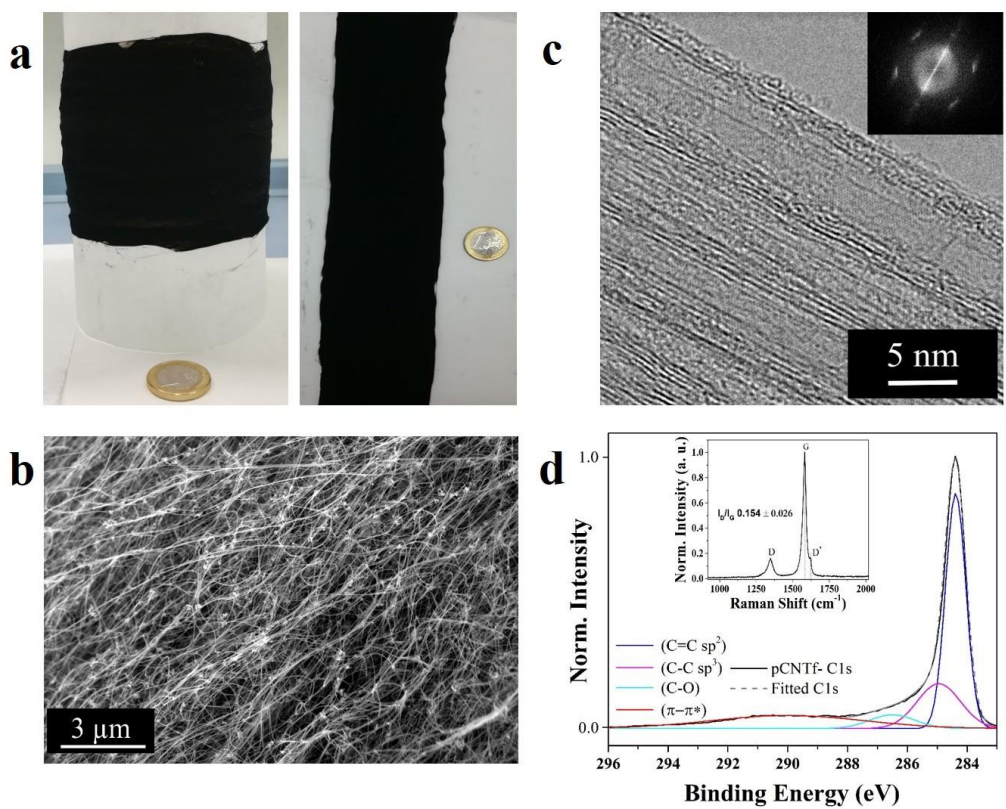

**Fig. 1** General morphology and composition of non-woven fabric of CNT fibres, showing their structure consisting of a porous network of crystalline domains (i.e. CNT bundles). a) Photograph of a typical electrode. b) SEM micrograph showing the network of CNT bundles. c) HRTEM image of highly graphitised CNT, with the Moiré pattern arising from CNT self-collapse and exhibiting hexagonal symmetry and armchair chiral angle (inset: fast Fourier transform). d) XPS spectra deconvolution analysis of CNT fibres with components corresponding to C=C ≈ 284.38 eV, C-C ≈ 284.95 eV, C-O ≈ 286.5 eV and π-π* transitions ≈ 290 eV. Inset includes the RAMAN spectra of the CNT yarns composed of low-defective CNTs.

The coexistence of mechanical robustness, high surface area and high conductivity is at the core of the extensive use of nanocarbon-based electrodes/current collectors in multiple energy-storing devices with augmented mechanical properties, ranging from flexiblity in bending [13] to stretchability [14] to semi-structural properties [15].

However, in spite of the strong technological interest in CNT fabrics and yarns as electrodes in energy-storage related processes, their structure remains poorly understood. There is generally a lack of accurate descriptors of the pore structure that can be linked to basic electrochemical properties, including those providing a better understanding of suspected volumetric changes upon liquid infiltration, [16] or under electrochemical stimuli. [17] Swelling and other forms structural re-arrangement are particularly relevant for these systems on account of the weak interaction between CNT bundles delimiting the pores. A method that could enable the characterisation of pores in nanocarbon-based electrodes during *operando* conditions in the presence of a liquid could provide valuable insights into structural changes during electrostatic ion adsorption or ion intercalation, amongst electrochemical processes. For general sample characterisation, such technique would also complement gas-adsorption textural measurements, which require relatively large CNT fibre volumes and can give substantial data scattering (gas-adsorption specific surface area (SSA) values for similar CNT fibres, for example, span from 75 to 515 $m^2 g^{-1}$). [13]

2D wide- and small-angle X-ray scattering has proven particularly useful to study the inherently irregular and hierarchical structure of CNT fibres, so far mainly to determine CNT alignment in oriented fibres because of its relation to longitudinal mechanical and transport properties. [18,19] There has been comparatively less progress on characterising the pore structure of these systems by SAXS. We recently demonstrated that CNT yarns present fractional Porod slopes and can be described as fractal structures with a surface fractal dimension of 2.5_2.8, depending on constituent CNTs. [20] Virtually all graphitic materials, from graphite to activated carbons to carbon black/polymer composites, have fractional Porod slopes and a fractal dimension of 2.3_2.9, [21-24] often across $10^6$ length-scales. [25]

A relevant, yet often overlooked feature of graphitic materials is the contribution of density fluctuations to SAXS data, widely studied by Ruland and co-workers in various carbon fibres and glassy carbon. [26,27] These density fluctuations arise from the distribution of interlayer spacings in graphite, effectively a microvoid system formed by the surface of the boundary between layers, leading to a scattering component scaling (in Kratky-smeared optics) with the reciprocal of scattering vector, $q$. This vast body of work showed that a correct structural analysis of graphitic ma-

terials by SAXS requires removing the term associated to density fluctuations from scattering intensity.[28]

Here, we show that pristine yarns of CNTs present such density fluctuations with considerable similarity to monolithic graphitic materials. Such observation also helps clarify the dimensional analysis that must be used to process SAXS data. This enables us to introduce a method to obtain structural parameters of CNT yarns from SAXS data, of the kind required to analyse electrochemical processes in traditional porous carbons *in situ*.[29,30] We demonstrate the application of the method by studying *in situ* electrochemical swelling of CNT yarn electrodes in an ionic liquid, a process whereby the specific surface of the fibre and specific capacitance increase irreversibly. The electrochemical conditions used here purposely avoid CNT functionalisation; thus, the combined WAXS/SAXS and electrochemical measurements shed light direct into the structure-capacitance, particularly in identifying the relevant length-scales involved in fibre swelling and the relation of this process to the quantum and Helmholtz components of total capacitance in these electrodes. As such, these findings have direct relevance to characterise electrodes based on nanocarbons for applications in energy storage, energy transfer, actuation and capacitive water purification, amongst others.

# 1 Results and Discussion

## 1.1 Structural parameters from SAXS analysis of pristine CNT yarn

The starting point of this study is to analyse a thin non-woven fabric of multiple CNT fibres. A typical CNT fibre electrode is produced by overlapping multiple individual CNT fibre filaments into a ($> 1000\ cm^2$) non-woven fabric (Fig. 1a) where the individual filaments essentially merge. Such material corresponds to the format of electrodes we have previously used in a wide range of applications in energy storage,[31,32] capacitive deionization,[33] solar energy conversion and electrocatalysis,[34] amongst others. It is also a similar format to sheets produced semi-industrially in scaled-up facilities.

As shown in the electron micrograph in Fig. 1b, the pore structure is highly irregular, with a hierarchical fibrous structure analogous to that found in a conventional yarn, such as cotton or wool.[35] The material can be visualised as network of elongated elements that overlap over substantial lengths but also branch out and thus give rise to porosity. The network consists of bundles of CNTs packed together in crystalline domains at a separation similar to turbostratic graphite. The CNTs have high aspect ratio (length/diameter: $10^6$) and long length ($\approx 1$ mm) and can thus form part of multiple bundles, conceptually similar to polymer chains in crystalline domains. As such, porosity is clearly independent of the degree of perfection of the CNT along the nanotube axis. Unlike conventional carbon-based electrodes, CNT fibre electrodes can be considered a macromolecular ensembles. Our standard synthesis conditions produce multi-walled CNTs with few layers (2–5 walls, MWCNTs) and $\approx 2$–6 nm diameter, which thus tend to self-collapse. An example of a HRTEM micrograph in Fig. 1c shows that nanotubes within bundles are substantially ordered, leading for example to well defined Moiré patterns (see inset Fig. 1c) arising from interference of the collapsed graphitic layers of CNTs with predominant armchair chirality. X-ray photoelectron spectroscopy (XPS) provides complementary evidence of the low defect density of the constituent CNTs. The C1s XPS spectrum included in Fig. 1d shows the strong predominance of $sp^2$-hybridized carbon (at $\approx 284.38$ eV), with a smaller component from $sp^3$-hybridised (at$\approx$285.1 eV) which includes double bond pyramidalisation.[36] Conjugation of the CNTs is also manifested by strong features at 290 eV and around 315 eV due to $\pi$-$\pi^*$ transitions and plasmons, respectively. Raman spectra show a low D/G ratio ($\approx 0.16$) and confirm the high order of the CNTs (see inset Fig. 1d).

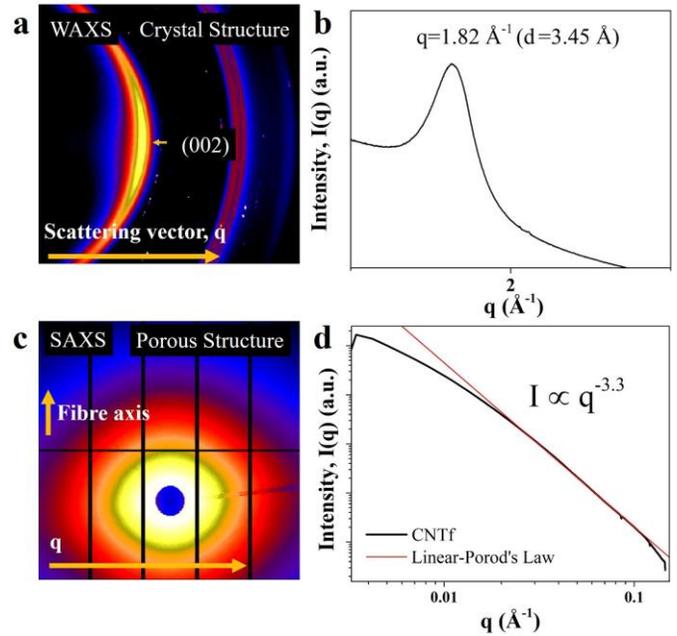

**Fig. 2** Typical WAXS/SAXS patterns and their respective radial profiles. a) WAXS pattern revealing the crystalline domains assigned to CNT bundles. b) Typical WAXS radial profile showing a broad (002) reflection due to a wide range of interlayer separations. c) 2D SAXS diffraction pattern due to porosity from imperfect packing of bundles. d) Typical SAXS radial profile showing scattering from the irregular pore structure comprising: a fractional Porod slope and a change in slope at very low q (q $\approx 0.014$ Å$^{-1}$) attributed to the sample mesoporosity.

2D WAXS/SAXS patterns capture the coexistence of crystalline domains and irregular pores in the network structure, as shown in the examples in Fig. 2 and their associated radial profiles after azimuthal integration. Scattering from graphitic layers in coherent domains produces the (002) reflection in the WAXS region, which is however, very broad and skewed towards lower q values ($1.815$ Å$^{-1}$) on account of the highly imperfect packing of CNTs, leading to an average interlayer spacing of $3.46$ Å. In the SAXS region the radial profile shows scattering indicative of an irregular pore structure dominated by the wide distribution of separation between bundles in the fibre. Plotted as log ($I_{obs}$) vs log (q) the SAXS data show a slopping profile and a constant linear component with Porod's slope of $-3.3$. At high $q$ values ($q > 0.08$ Å$^{-1}$) there is a small peak associated with the form factor of the

CNTs,[37,38] which can be more clearly resolved in the equatorial component of highly oriented CNT fibre samples (the ones used in this work are purposely synthesised with nearly isotropic structure in the fibre/fabric direction). At very low $q$ values ($< 0.014$ Å$^{-1}$) there is a change in slope, which we attribute to contributions from the mesoporous structure. For reference, $N_2$ measurements indicate a pore distribution starting from mesopores around 20 nm and extending to the macropore range.[20]

The fractional Porod slope is typical of graphitic materials,[37] and taken as indicative of a fractal (rough) surface. In this case, it leads to a surface fractal dimension of 2.7, which is in agreement with values obtained from $N_2$ adsorption measurements[20] and in general with values for many different graphitic carbons, 2.2 − 2.8,[39,40] also determined by dynamic light scattering[41] or SEM[42]. This fractional Porod slope is a consequence of the scattering intensity having a contribution from density fluctuations. It produces a one dimensional term of scattering intensity, which in pin-hole optics scales with $q^{-2}$. This term can be observed as a linear region in a plot of $I_{obs}(q)\, q^4\, vs\, q^2$, as shown in Fig. 3a, and the data can be corrected accordingly $I_{corr}$ ($I_{corr} = I_{obs} - bq^{-2}$) for further analysis.

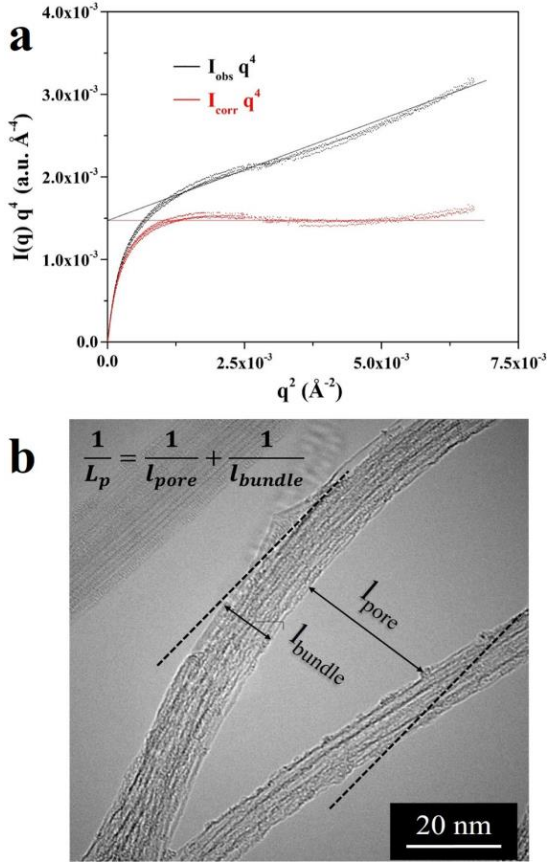

**Fig. 3** Scattering component arising from density fluctuations. (a) A plot of $I(q)\, q^4\, vs\, q^2$ shows a linear region, corresponding to density fluctuations.(b) Main pore structure descriptors of CNT fibres.

These density fluctuations, proposed and widely analysed by Ruland,[27,28,43] correspond to fluctuation in electronic density in graphitic systems arising from variations in size and shape of the graphitic layer stacks with respect to Bernal graphite ($d_{002, graphite}$ = 3.35 Å$^{-1}$). As such, they are present in virtually all materials made up of graphitic domains irrespective of their pore structure or degree of crystallinity, including glassy carbon,[26] carbon fibres from different precursors,[27,43,44], nanoporous activated carbons,[45] non-graphitizable carbons[46] and now identified in macroscopic CNT fibres. We note, however, that in traditional carbons graphitisation treatments increase crystal size both *parallel* and *perpendicular* to the basal plane, thus reducing the contribution from density fluctuations,[27,46] whereas in the case of CNT aggregates these two terms are independent. In general, the large anisotropy of graphitic layers and their high in-plane stiffness can hinder commensurate stacking and imply that stacking faults propagate over long lengths. In the case of CNTs, curvature and differences in chiral vector make commensurate stacking even less likely.

### 1.2 Pore structure model

Equipped with adequate correction of the data to account for density fluctuations, it is possible to calculate descriptors of the pore structure, some of which are shown schematically in Fig. 3b superimposed onto a HRTEM micrograph of a typical pore opening. Assuming a three-dimensional two-phase system with sharp boundaries, the elongated scattering elements in the CNT yarn can be characterised in terms of a coherence length ($l_c$) and the Porod (average chord) length ($L_p$).[47]

$$l_c = \frac{\int_0^\infty qI(q)dq}{\int_0^\infty q^2 I(q)dq} \qquad (1)$$

$l_c$ corresponds to the average lateral distance over which the correlation length is present. It is sensitive to electron density fluctuations at the periphery of the voids.

$L_p$ is the average lateral lenght of scattering elements along the sample cross section, given by

$$L_p = \frac{Q}{K_p} \qquad (2)$$

$Q$ is the invariant, equivalent to the scattering power ($k$) per volume probed ($V$) and experimentally determined as

$$Q = k/V = \int_0^\infty q^2 I(q) dq \qquad (3)$$

Through its dependence on the volume fraction of the two phases in the system (i.e. CNT and air) and the difference in electron density (($\Delta\rho)^2$), it provides a measure of sample porosity $P$.

$$Q = 8\pi(\Delta\rho)^2 P(1-P) \qquad (4)$$

$K_p$ is the Porod's constant, obtained experimentally as

$$K_p = \lim_{q \to \infty} q^4 I(q) \qquad (5)$$

And which is related to the surface to volume ratio $S_V$

$$K_p = 2\pi(\Delta\rho)^2 S_V \qquad (6)$$

**Table 1** Structural parameters derived from SAXS data analysis after correction of intensity by subtraction of density fluctuations (data without correction in parenthesis)

| Carbonaceous Material | $l_c$ (Å) | $L_p$ (Å) | DFs | $l_{pore}$ (Å) | $l_{bundle}$* (Å) | $SSA_{SAXS}$ ($m^2g^{-1}$) | P |
|---|---|---|---|---|---|---|---|
| MWCNTs fibre (this work) | 53 (60) | 106 (91) | 1.913 | 305 (255) | 161 (142) | 259 (305) | 0.66 (0.64) |
| SWCNTs fibre (this work) | 58 | 62 | 0.05 | 205 | 89 | 656 | 0.70 |
| CF (PAN, pitch based) ([27,43,44]) | 9.17-25.24 | 6.18-13.88 | ≈ 0-0.012 | 6.85-17.75 | 36.70-124 | 125-367 | 0.097-0.33 |
| Activated Carbon Cloth ([45]) | 12 | 5.4 | 0.25 | 7.6 | 19 | 1090 | 0.29 |
| Porous Carbon ([48]) | – | 17-22 | – | 16.69-115.47 | 20.98-31.88 | 529-915 | 0.69-0.81 |
| Carbon Black ([49]) | 276-307 | 175-186 | – | – | – | 78.7-89.5 | 0.67-0.71 |
| Activated Carbon ([21]) | – | 6.6-6.8 | – | 12.3-12.8 | 13.5-15.2 | 725-750 | 0.45-0.49 |

*$l_{bundle}$ or $l_{matter}$

From equation 6 it is possible to obtain the surface area per unit mass, i.e. the specific surface area (*SSA*) by taking the ratio of $S_V$ over the material's density, and compare with gas adsorption measurements, for example. In such case, the density used for normalisation must be the apparent density; for the samples in this work close to 0.33 $gcm^{-3}$.

$L_p$ is related to the lateral size of scattering elements and thus to average pore size $l_{pore}$, average bundle lateral size, $l_{bundle}$ (Fig. 3b) and porosity

$$L_p = l_{pore}(1 - P) = l_{bundle}P \qquad (7)$$

$$1/L_p = 1/l_{pore} + 1/l_{bundle} \qquad (8)$$

Finally, the magnitude of density fluctuations can be calculated based on the expression previously used to describe similar needle-shaped pores in various CFs. [27,43,46]

$$DFs = \frac{8\pi^2 b_2 P}{a_3 Q} = \frac{\langle \Delta^2 a_3 \rangle}{\langle a_3 \rangle^2} + \frac{\langle \Delta^2 L \rangle}{\langle L \rangle^2} \qquad (9)$$

Where $b_2$ is extracted from the slope of the Ruland plot ($I(q)q^4$ vs. $q^2$) (Fig. 3a) and $a_3$, the distance between graphitic layers in Bernal graphite ($d_{002}$). Equation 9 also relates the density fluctuations to two different components: one arises from fluctuations in interlayer spacing $\frac{\langle \Delta^2 a_3 \rangle}{\langle a_3 \rangle^2}$ and another from fluctuations in stack size $\frac{\langle \Delta^2 L \rangle}{\langle L \rangle^2}$, although their contributions cannot be separated using only SAXS data.

The values for structural parameters obtained by SAXS analysis on fibre made up of either MWCNTs or SWCNTs, are presented in Table 1, including for comparison literature data for related carbonaceous materials used in energy storage and other applications. Firstly, for MWNTs we note the large effect of DFs on the various structural parameters (see values without DF correction in parenthesis), which highlights the importance of the corrections introduced in this work. The DFs are intrinsically high in this system, roughly a factor of 100 higher than for CF, but again, only reflecting imperfect staking of graphitic layers. In contrast, the density fluctuations for the SWCNT fibre are very small. Indeed, in SWNTs fibres there is insignificant stacking of graphitic layers, with an extremely weak (002) peak in WAXS. This observation helps establish that the density fluctuations observed in SAXS from MWCNTs fibres arise predominantly from intra-tube scattering.

Overall, the corrected structural parameters derived from SAXS are in agreement with the mesoporous internal structure of MWCNTs fibres probed by gas-adsorption and other methods. The pores and bundles have average sizes of 305 Å and 161 Å, respectively. Taking the bundle size as equivalent to the stack size, the value for MWCNTss fibres is in the same range than that of CF (36.7 – 124Å). $SSA_{SAXS}$ comes out as 259 $m^2g^{-1}$, which is close to the one obtained from $N_2$ gas adsorption, [9] 250 $m^2g^{-1}$. Similarly, porosity is 0.66, which is comparable to the porosity normally shown in activated carbons and in carbon black materials and noticeable higher than that of CF. But a key aspect for the correct interpretation of these values is to observe that both SSA and P determined from SAXS are relative to Bernal graphite. [43] For CFs, their extremely high $SSA_{SAXS}$ from 125 to 367 $m^2g^{-1}$ contrasts with a very low SSAs around 0.4 $m^2g^{-1}$ for closed or micro-pores inaccessible to the gas. Indeed, porosity for CF is similar by SAXS (0.326) and He adsorption (0.327), indicating that SAXS probes almost all pores. [43] In the case of CNT fibres $SSA_{SAXS}$ has a component from "closed porosity", for instance the inside of the CNTs, hence the substantial differences in $SSA_{SAXS}$ values for MWCNTs and SWCNT fibres in Table 1. It is possible to independently calculate porosity from apparent density (0.329 $gcm^{-3}$) and theoretical density $P = 1 - \rho_{app}/\rho_{th}$. For SAXS, the correct theoretical density is with respect to a stack of graphitic planes, given by

$$\rho_{CNTyarnTh} = 2.268 \frac{d_{002graphite}}{d_{002CNTfibre}} \qquad (10)$$

Where 2.268 $gcm^{-3}$ is the density of single-crystal graphite [44] and $d_{002}$ the interlayer spacing. This gives a porosity of 0.85, which is higher than the value in Table 1. For reference, taking instead, the theoretical density of a bundle of SWCNTs of identical diameter, $\rho_{th} \approx 1.8$ and $P = 0.82$. Ultimately, the "correct" value of $P$ depends on the process of interest. More importantly, the results presented above show that for CNT fibres, and probably also related nanocarbon ensembles, the predominance of mesoporosity makes SAXS ideally suited to study ion-electroadsorption and similar interfacial processes. As an example, below we present results from *in situ* and *ex situ* SAXS measurements, complemented with WAXS analysis, during electrochemical swelling of MWCNTs fibres. Whereas WAXS/SAXS readily provide information about the electrode structure *in situ*, we have found inherent limitations in using BET gas-adsorption measurements to characterise these materials (ESI).

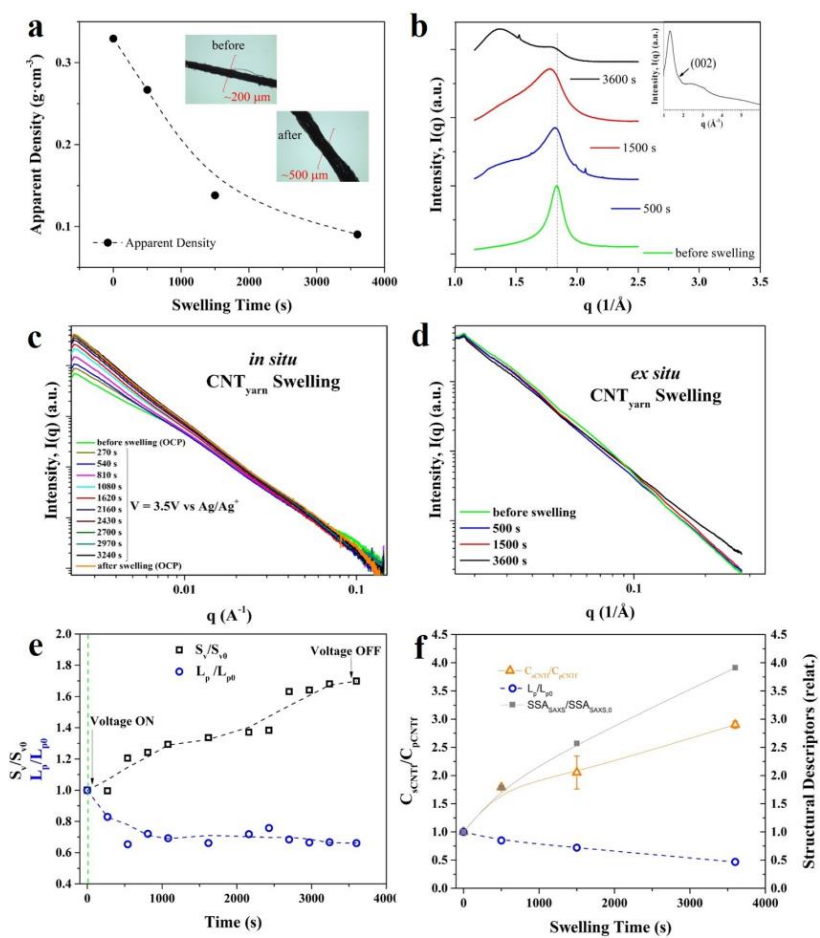

**Fig. 4** *In situ* WAXS/SAXS study of the electrochemical swelling of CNT yarns at 3.5 V. (a) Apparent density decrease (increase in CNT fibre diameter) as a consequence of the swelling duration. Inset: Optical micrographs showing direct evidence of swelling after the test. (b) WAXS radial profiles of clean (IL removal) and dry pristine and swollen CNT fibres at different swelling times, indicating a shift in the (002) reflection position. Inset: *in situ* WAXS radial profiles give limited information due to strong contribution from the IL. (c) SAXS radial profiles during *in situ* the electrochemical test showing an increase in intensity at low q values. (d) SAXS radial profiles at different swelling times *ex situ* showing an increase in intensity at high q values. (Note different q ranges) (e) Changes of structural descriptors during *ex situ* electrochemical swelling. (f) Correlation between the modification of structural descriptors (i.e. $SSA_{SAXS}$, $L_p$) and the increase in the electrochemical capacitance of samples treated at different swelling times.

### 1.3 Monitoring structural changes by SAXS during electrochemical swelling of CNT yarns

Electrochemical swelling of CNT fibres was induced by applying a constant electrochemical potential of -3.5 V vs Ag/$Ag^+$ inside a custom built three-electrode cell with ionic liquid. 1-Butyl- 1-methylpyrrolidinium (bis(trifluoromethanesulfonyl)imide, $PYR_{14}TFSI$, was chosen as inert electrolyte exclusively because of its high chemical stability (3.5 V in a device,[9,50] 6.1 V according to supplier). Amongst the myriad of methods developed to modify carbonaceous electrodes,[51,52] the conditions in this work were chosen mainly on the basis that they preserve electrode electrical conductivity and do not appreciably alter surface chemistry nor introduce pseudocapacitive reactions, thus providing the opportunity to unambiguously relate X-ray-derived pore structure parameters to electrochemical capacitance changes.

A general indication of the effectiveness of the swelling process is gained by direct optical observation of the large increase in sample volume and reduction in apparent volumetric density upon swelling (Fig. 4a) WAXS measurements were performed *in situ* during the electrochemical test, but the strong overlap between the structure factor of the IL and the (002) reflection from CNT fibres impedes accurate monitoring of the process *in situ* (see inset Fig. 4b). However, *ex situ* WAXS shows a progressively broader (002) reflection skewed towards low q values as a result of increased separation between bundles (Fig. 4b).

SAXS measurements were carried out both *ex* and *in situ* in order to demonstrate the correlation between structural changes and capacitance under different experimental conditions. Note that in *in situ* measurements the electrode is fully infiltrated with ionic liquid, whereas in *ex situ* it has been cleaned and dried. For the *in situ* experiment, measurements were conducted at open circuit potential (OCP), during the swelling process and at OCP immediately after swelling. The evolution of SAXS radial profile is plotted in Fig. 4c. Overall, the increase in the scattered intensity confirms the formation of new interface boundaries between CNT and the electrolyte (ionic concentration changes in pure IL can be discarded). At high q values there is a substantial change in the slope of the radial profile, but which is more visible in the *ex situ* data collected with a larger q-range (Fig. 4d).

The interest then, is in quantifying the increase in surface area and identifying the structural elements or length-scales relevant in the swelling process. This information from SAXS is summarised in Fig. 4e,f, presented as the changes in surface to volume ratio ($S_V/S_{V0}$) and Porod length ($L_p/L_{p0}$) during electrochemical swelling, calculated according to equations 6 and 2. $S_V/S_{V0}$ shows a large increase during the swelling. This change in surface area remains constant after the test is finished and the sample is at OCP, which confirms that swelling is irreversible. $L_p$ shows an analogous trend but of opposite sign, indicating that as SSA increases the average distance between bundles decreases. Very importantly, capacitance measurements at different swelling times show a good correlation with the increase in surface area obtained from SAXS (Fig. 4f). The direct correlation between the modification of the MWCNTs fibre structure and the specific capacitance supports the view that swelling produces the creation of new electrode/electrolyte interfaces due to branching of bundles, which increases the number of sites available for ion electrosorption and thus specific capacitance. We note that the absolute change in $SSA_{SAXS}$ obtained from *ex situ* data in Figure 4d is more accurate due to a higher scattering intensity and larger q-range as a result of experimental conditions used for such measurements.

Pristine and swollen MWCNTs-based fibre electrodes were further analysed by electrochemical impedance spectroscopy (Fig. 5a and Fig. S1, ESI) and cyclic voltammetry (Fig. 5b) in a three electrode cell using $PYR_{14}TFSI$ electrolyte. Fig. 5a presents the real part of capacitance ($C'$) against frequency, determined from EIS and which reflects features of the dominant energy storage mechanism.[53] Both before and after swelling the samples show a plateau at low frequencies, typical of the EDL formation in mesoporous materials without diffusion limitations. EIS measurements at different swelling states additionally show that equivalent series resistance (ESR) remains fairly constant (see Nyquist plot in Fig. S1, Table S1 and Fig. S2a ESI). This confirms that the dominant process is swelling by separation of adjacent bundles, rather than extensive individualisation (i.e. exfoliation) of nanotubes within bundles which would substantially increase electrical resistance. As expected from micromechanical considerations, the weakest interface in CNT fibres is that between bundles, rather than between CNTs in a bundle, and thus the most likely scenario is that swelling is mainly produced by ions separating bundles.

In agreement with these data, cyclic voltammograms, CVs, (Fig.5b) of MWCNTs fibres also show similar lineshapes before and after swelling, albeit with a larger area in the swollen sample (see CVs of SWCNTs samples, Fig. S2b in ESI). In addition, we note that the swollen MWCNTs electrodes exhibit coulombic efficiency higher than 95% after 5000 cycles (see Fig. S3, ESI) and have no indication of redox peaks.

These results are in line with the structural changes of the electrodes observed by SAXS, confirming that the porosity increase in swollen CNT fibres does not come from the opening of micropores (e.g. the inside of the CNTs), pseudocapacitive reactions or intercalation. More importantly, the implication is that SAXS can be effectively used to analyse electrochemical processes in mesoporous nanocarbon ensembles. An example is the contribution of low-dimensional properties of nanocarbons to the bulk electrochemical properties of their macroscopic ensembles.

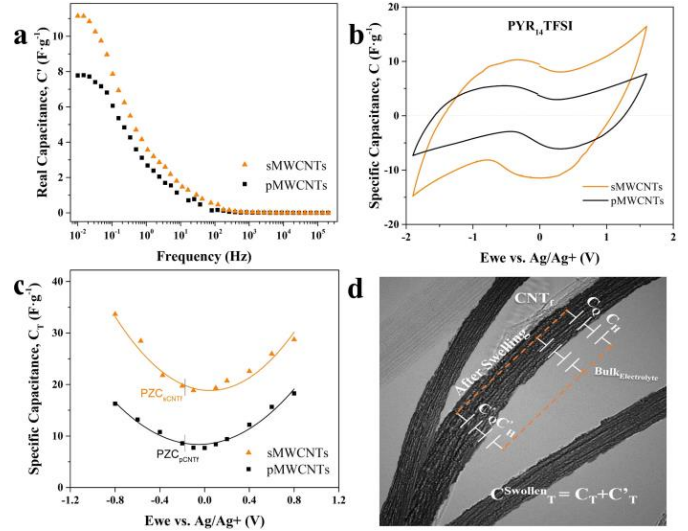

**Fig. 5** (a) Dependence on frequency of real capacitance (C') for the pristine (pMWCNTs) and swollen (sMWCNTs) fibre electrodes. (b) CVs of the pMWCNTs and sMWCNTs in $PYR_{14}TFSI$ at 20 $mVs^{-1}$ and 3.5 V operating window (c) V-plot comparison between pMWCNTs and sMWCNTs electrodes, including the potential at zero charge (PZC), calculated from electrochemical impedance at 10 mHz and different bias voltage. (d) HRTEM-schematic illustration of the Density-of-States (*DOS*) in the *CNTs bundle*, the charge distribution at the bulk electrolyte/electrode interface and therefore, the *quantum* capacitance ($C_q$), the *Helmholtz* capacitance ($C_H$) and the sum *total* capacitance ($C_T$).

After swelling, the CV still shows the distinctive butterfly-shape for CNTs ensembles due to their small quantum capacitance [9,54] and small Debye ionic screening length reminiscent of the low density of states (DOS) of the constituent CNTs.[55] In this context, a plot of EIS-determined specific capacitance ($C_T$) against electrochemical potential is particularly instructive because its lineshape has close correspondence to the DOS when the Helmholtz ($C_H$) and quantum capacitance ($C_q$) contributions are comparable (Note that they are in series and therefore the smaller one dominates). Indeed, a comparison of the experimental $C_T$ with the theoretical $C_q$ of a superposition of individual single-walled CNTs (see Fig. S4, ESI) shows a pleasant similarity and suggests that the quadratic shape of $C_q$ is reflecting the low DOS of CNTs near the Fermi level and higher DOS close to energies corresponding to Van Hove singularities. However, in the case of CNT fibre electrodes, the joint density of states (*JDOS*) corresponds to the bundles, which is not a direct superposition of those of individual CNTs because of electronic coupling and other forms of overlap of their electronic structures,[56,57] and thus the comparison must be taken with caution. The relevant result is that electrochemical swelling mainly moves the graph upwards without affecting its lineshape with respect to electrochemical potential (Fig. 5c). This, together with the structural analysis discussed above, suggests again that there is not substantial individualization of CNTs, and instead swelling produces separation of bundles. Thus, bundles are still held firmly in crystalline regions. This is in agreement with theoretical calculations showing that individual CNTs have around 40 times lower $C_q$ that their corresponding bundles, which implies that individualising (exfoliating) CNTs from a bundle would augment its contribution to $C_T$ by about two orders of magnitude. Clearly this is not the case.

Finally, we present a schematic (Fig. 5d) with the aid of a HRTEM micrograph of a pore opening that illustrates the key features of the swelling process. The combination of *in situ* SAXS measurements, and electrochemical measurements after swelling point to electro-adsorbed ions acting as a wedge that opens a bundle into smaller ones, but still large enough that the contribution from $C_q$ and $C_H$ are similar. This is equivalent to the newly-exposed surface producing more elements in parallel with those previously available for EDL formation (see equivalent circuit in Fig. 5d).

## 2 Experimental

### 2.1 Materials synthesis

Highly graphitised CNT yarns were synthesized continuously by the direct spinning chemical vapor deposition method from a vertical reactor at a winding rate set at 5 $mmin^{-1}$, the source of carbon was butanol, ferrocene was used as iron catalyst and thiophene as promoter. The concentration of precursors butanol:thiophene:ferrocene ($97.7:1.5:0.8$) was chosen to produce CNTs predominantly made up of $3-5$ layers (multiwall CNTs, MWCNTs) with average diameter of $2.5\text{-}5$ nm, and ($99:0.2:0.8$) to produce SWCNTs of diameter around $1.5$ nm.[58]

### 2.2 Materials Characterization

SEM and TEM analysis were carried out with a FIB-FEGSEM Helios NanoLab 600i (FEI) at 15 kV and a JEOL JEM 3000F TEM at 300kV, respectively.

Simultaneous 2D WAXS/SAXS patterns of CNT yarns samples were collected at ALBA synchrotron facilities (Barcelona, Spain) at NCD-Beamline 11. The radiation length was 1 Å and a microfocus spot diameter of 10-microns, approximately, was used. Samples consisted in yarns of multifilaments, typically of $\approx 1$ mm thickness. Scattering of the samples was acquired in the range of $0.0035 < q$ (Å$^{-1}$) $< 0.15$. Sample holder positions are calibrated using a reference material: silver behenate (AgBh) and chromium oxide ($Cr_2O_3$) for SAXS and WAXS measurements, respectively. Data were corrected for background scattering. SAXS radial profiles were obtained after full $360^o$ integration. Figures of merit have been calibrated into units of scattering cross section, i.e. scattering intensities are given by absolute units normalized by probed volume irradiated by the x-ray beam, i.e. Å$^{-3}$. In this work, $I$ and $I_{obs}$ refer to scattering intensities with and without density fluctuations corrections, respectively. WAXS/SAXS data treatment is discussed in details in the *Results* section.

Nitrogen ($N_2$) adsorption at 77K of CNT fibres was measured with Quantachrome Instrument (Quadrasorb SI, version 5.03). Samples were previously degassed at 150 $^o$C under vacuum for 20 h. The specific surface areas were calculated by using B.E.T equation and Quadrachrome *ASiQwin$^{TM}$* software. Results are discussed in ESI.

### 2.3 Electrochemical Characterization and Swelling Tests

Samples for electrochemical swelling consisted of arrays of multiple MWCNTs fibre filaments, similar to a yarn or a tow, with length of 1 - 6 cm (depending on the electrochemical cell used), diameter around 1 mm and a mass loading of 2 - 4 $mgcm^{-2}$. As-formed MWCNTs fibres were shaped into a yarn and placed between Mica windows within the electrochemical cell and then, filling with *PYR$_{14}$TFSI* to carry out an appropriate impregnation of MWCNTs fibres with this electrolyte before swelling.

The solvent-induced swelling process consists in the adsorption of *PYR$_{14}$TFSI* ionic liquid (bis(trifluoromethanesulfonyl)imide, >99.5% Solvionic) between the carbon structures of the MWCNTs network under chronoamperometry (CA) conditions (−3.5 V for 60 minutes *in situ* WAXS/SAXS measurements or different swelling times, *ex situ* study) in a three-electrode configuration. IL was chosen as an inert electrolyte to carry out the swelling due to its high chemical stability (3.5 V in a device,[9,50] 6.1 V according to supplier, *Solvionic*). A pristine MWCNTs yarn ($\approx 2.5$ mg of active material mass) was set as working electrode, platinum mesh as counter electrode and a silver wire was used as a pseudo reference electrode.

Pristine (pMWCNTs, pSWCNTs) and swollen (sMWCNTs,sSWCNTs) CNTs fibres were electrochemically characterised based on cyclic voltammetries (CV) and electrochemical impedance spectroscopy (EIS) measurements to obtain specific capacitance ($C_s$) and Nyquist plot, respectively. CVs at a scan rate of 20 $mVs^{-1}$ in *PYR$_{14}$TFSI* were performed within an operating voltage window of 3.5 V. $C_s$ was calculated by integrating area under the anodic and cathodic curves and then, dividing by active material mass.

Measurements of capacitance in polarized conditions, before and after MWCNTs yarn swelling, were performed to determine potential of zero charge (PZC) from EIS at 10 mHz and different bias voltage, following methodology fully described before.[9]

All electrochemical tests (CVs, CA and EIS) including WAXS/SAXS *in situ* measurements were carried out in the three-electrode configuration using a Biologic SP-300 potentiostat-galvanostat coupled with EC-Lab v11.20 software.

## 3 Conclusions

A central characteristic of CNT fibre electrodes and related nanocarbon ensembles is their complex mesoporous structure, resulting from imperfect packing of highly conjugated and long CNTs. SAXS and WAXS are ideal techniques to probe this highly-crystalline, yet porous and hierarchical structure. SAXS shows a deviation from Porod's law, which arises due to density fluctuations of the stacks of graphitic domains. This work shows that it is critical to subtract these fluctuations from the observed scattering intensity in order to obtain accurate structural descriptors for fibres made up of CNTs with several layers, with SWCNTs fibres showing a much lower contribution. After such correction, the SSA determined by SAXS is remarkably similar to that observed by gas-adsorption. This provides very strong support for the analytical treatment of the data proposed here, while also highlighting the inherently predominant open porosity in nanocarbon ensembles, effectively molecular solids.

The density fluctuations observed in CNTs fibres have been widely reported in other graphitic systems, such as carbon fibres. However, whereas in CF and traditional carbons stack size parallel and perpendicular are strongly related and compounded in the degree of graphitisation of the material, in CNTs fibres molecular conjugation and stack formation are in principle independently fixed at the points of synthesis and assembly, respectively. As such, CNTs fibre longitudinal fibre properties are likely to be linked to density fluctuations by virtue of their relation to inter-bundle stress and charge transfer processes. Current efforts are directed at exploring such possible correlation between density fluctuations and longitudinal mechanical and electrical properties.

Electrochemical charging of a CNTs fibre in ionic liquids produces swelling of the CNTs fibres and roughly doubling of capacitance, without significant loss of electrical conductivity. *In situ* WAXS/SAXS measurements during the process show that the process is dominated by opening of bundles, but without individualisation of the CNTs. The increase in surface area determined by SAXS both *in situ* and *ex situ* are in agreement with the increase in capacitance determined independently in an electrochemical cell. This supports the use of WAXS/SAXS for the study of CNTs fibre and related nanocarbon electrodes in various other electrochemical processes, such as in electrochemical actuation and in batteries. Work is in progress to relate fibre morphology (e.g. CNTs type, degree of alignment) to the changes in electrode properties under different electrochemical stimuli.

Finally, combined electrochemical and WAXS/SAXS measurements during swelling show no major changes in the contributions to the total capacitance from the Helmholtz and quantum capacitances. According to these results, altering this balance to access predominantly the quantum capacitance and thus the JDOS will require some degree of individualization of the CNTs.